\documentclass[aps,pra,twocolumn,groupedaddress]{revtex4}

\def\kv{{\bf k}}
\def\rv{{\bf r}}
\def\Rv{{\bf R}}

\begin{document}
\title{What is a crystal?}
\author{Ron Lifshitz}
\affiliation{Condensed Matter Physics 114-36, California Institute of
  Technology, Pasadena CA 91125, U.S.A.}
\thanks{Permanent address: School of Physics \& Astronomy, Raymond and
  Beverly Sackler Faculty of Exact Sciences, Tel Aviv University,
  69978 Tel Aviv, Israel}
\email{ronlif@tau.ac.il}

\date{January 1, 2007}

\begin{abstract}
  Almost 25 years have passed since Shechtman discovered
  quasicrystals, and 15 years since the Commission on Aperiodic
  Crystals of the International Union of Crystallography put forth a
  provisional definition of the term {\it crystal\/} to mean ``any
  solid having an essentially discrete diffraction diagram.'' Have we
  learned enough about crystallinity in the last 25 years, or do we
  need more time to explore additional physical systems? There is much
  confusion and contradiction in the literature in using the term
  crystal. Are we ready now to propose a permanent definition for {\it
    crystal\/} to be used by all? I argue that time has come to put a
  sense of {\it order\/} in all the confusion.
\end{abstract}


\maketitle

\begin{quote}
{\it Confusion is a word we have invented for an order which is not
  understood.}
\begin{flushright}Henry Miller\end{flushright}
\end{quote}

\section{Periodicity or order?}

For almost two millennia crystallography was predominantly concerned
with the external morphology of crystals. Crystallographers studied
the naturally-occurring facets of crystals, which always intersect at
precise and characteristic angles. It was only in the 17$^{th}$
century that modern crystallography was born, thanks to the brilliant
idea---attributed to such great scientists as Kepler and Hooke---that
crystal shapes were the result of internal {\it order\/} of ``atomic''
units. 

In his study of tilings of the plane by polygons, Kepler~\cite{kepler}
was quick to realize that very few regular polygons---namely the
triangle, the square, and the hexagon---can tile the plane without
introducing overlaps or leaving holes. Yet, this observation did not
deter him from constructing a well-ordered tiling of the plane,
consisting of pentagons and decagons that requires some of the
decagons to overlap, and leaves holes in the form of 5-fold stars---a
tiling that was to be rediscovered by Penrose more than 350 years
later~\cite{penrose,gardner}. Without realizing it, Kepler had
discovered some of the basic properties of aperiodic order. In his
words:
\begin{quote}{\it ``If you really wish to continue the pattern,
    certain irregularities must be admitted, two decagons must be
    combined, two sides being removed from each of them. As the
    pattern is continued outwards five-cornered forms appear
    repeatedly: \dots So as it progresses this five-cornered pattern
    continually introduces something new.  The structure is very
    elaborate and intricate.''}\end{quote}

Unfortunately, by the end of the 18$^{th}$ century, when Ha\"uy began
formulating the mathematical theory of crystallography~\cite{hauy},
Kepler's insightful drawings of aperiodic tilings with decagonal
symmetry were long forgotten. Mathematical crystallography was founded
upon the premise that the internal order of crystals was necessarily
achieved through a periodic filling of space.  Thus, crystallography
treated {\it order\/} and {\it periodicity\/} synonymously, using
either property interchangeably to define the notion of a {\it
  crystal.}  The periodic nature of crystals was ``confirmed'' with
the discovery of x-ray crystallography and numerous other experimental
techniques throughout the 20$^{th}$ century. Being more ``elaborate
and intricate'' and less commonly found in Nature, aperiodic crystals
were completely overlooked. Periodicity became the underlying
paradigm, not only for crystallography itself, but also for other
disciplines such as materials science and condensed matter physics,
whose most basic tools, like the Brillouin zone, rely on its
existence.

This historical oversight was corrected with Shechtman's dramatic
discovery of quasicrystals in 1982~\cite{shechtman}---a discovery that
sparked a {\it bona fide\/} Kuhnian scientific
revolution~\cite{cahn,rebirth}.  In his description of the event,
Hargittai~\cite{hargittai} comments that
\begin{quote}{\it``\dots the discovery was a kind of legalistic discovery.
    This happens when the human classification system is more
    restrictive than the laws of nature and discoveries appear to
    break the laws that had been artificially constructed in the first
    place.''}\end{quote} Nature had found a way of achieving order
without periodicity, and Shechtman was careful enough not to dismiss
it as an experimental artifact, as many must have done before him. He
confronted a skeptical scientific community that was unwilling to
relinquish its most basic paradigm that order stems from periodicity.
In an article, suggesting a description of Shechtman's icosahedral
quasicrystal as multiple twinning of periodic cubic crystals---a
description that was later shown to be
incorrect~\cite{bancel}---Pauling~\cite{pauling} concluded by saying
that\begin{quote}{\it ``Crystallographers can now cease to worry that
    the validity of one of the accepted bases of their science has been
    questioned''}.\end{quote}

Today, thousands of diffraction diagrams later, compounded by
high-quality experimental data---such as images from high-resolution
transmission electron microscopes and atomic-resolution scanning
tunneling microscopes---the existence of order without periodicity has
been unequivocally established. Not only has the periodicity paradigm
been questioned, as Pauling worried, it has been completely
invalidated. It is time now, as Ben-Abraham~\cite{siba} argues in this
volume, to make a clear statement associating the notion of a {\it
  crystal\/} with its more fundamental and underlying property of {\it
  order,} rather than the property of {\it periodicity,} which is only
one way of achieving order (even if it is the more common way in which
Nature chooses to do so). According to this, crystals that are
periodic should explicitly be called {\it periodic crystals,} all
others should be called {\it aperiodic crystals.}

Some may object by saying that there are no right or wrong
definitions, and that a definition is merely a matter of mutual
agreement. They would further claim that the notion of a crystal
should strictly be associated with periodicity, because over a century
of literature in mathematical crystallography contains statements and
proofs about crystals that are, in fact, true only for periodic
crystals. A change of the definition would cause great confusion when
reading this literature. This type of argument is typical of the
paradigm-shifting stage we are currently undergoing in this scientific
revolution~\cite{kuhn}.  Nevertheless, the discovery of quasicrystals
demands such a paradigm shift and along with it a new definition of
the notion of a crystal.  One simply needs to be aware, whenever
reading old literature, that the term crystal may be used in its old
restricted meaning.


\section{But, what is order?}

In 1992, the International Union of Crystallography, through its
Commission on Aperiodic Crystals~\cite{iucr}, published a provisional
definition of the term {\it crystal\/} that abolishes {\it
  periodicity,} but does not go as far as naming {\it order\/} as its
replacement. The commission was not ready to give precise microscopic
descriptions of all the ways in which order can be achieved. Clearly,
periodicity is one way of achieving order, quasiperiodicity as in the
Penrose-Kepler tiling is another, but can we be certain that there are
no other ways that are yet to be discovered?  The Commission opted to
shift the definition from a microscopic description of the crystal to
a property of the data collected in a diffraction experiment. It
decided on a temporary working-definition whereby a {\it crystal\/} is
\begin{quote}{\it ``any solid having an essentially discrete
    diffraction diagram.''}\end{quote} The new definition was left
sufficiently vague so as not to impose unnecessary constraints until a
better understanding of crystallinity emerges. 

The 1992 definition is consistent with the notion of long-range
order---one of the basic notions of condensed-matter
physics~\cite{anderson,sethna}---dating back to ideas of Landau in
which the symmetry-breaking transition from a disordered
(high-symmetry) phase to an ordered (low-symmetry) phase is quantified
by the appearance of a non-zero {\it order parameter}. Stated in plain
words, long-range order---or in the context of our current discussion,
long-range positional order---is a measure of the correlations between
the positions of atoms in distant regions of the material. It measures
the extent of our ability to describe the positions of atoms in far
away regions of space, based on our knowledge of their positions
nearby.

How is this related to the requirement of having an essentially
discrete diffraction diagram? If $\rho(\rv)$ is a density function,
describing a certain material, then the function that measures
correlations between two points, separated in space by a vector
$\Rv$---the two-point autocorrelation function---is known in
crystallography as the {\it Patterson function,}
\begin{equation}\label{patterson}
P(\Rv)=\lim_{V\to\infty} {1\over V}\int d\rv
\rho(\rv)\rho(\rv-\Rv).
\end{equation}
Its Fourier transform is exactly the intensity $I(\kv)=|\rho(\kv)|^2$
that is measured in most diffraction experiments. It was realized long
ago ({\it e.g.} Ref.~\cite{mermin}), that the order parameter that
signals a transition from a disordered liquid to an ordered solid,
indicating the emergence of non-trivial correlations as measured by
the Patterson function~(\ref{patterson}), appears in the form of delta
functions, or {\it Bragg peaks,} at non-zero wave vectors in the
diffraction diagram. Thus, the observation of essentially discrete
points in the diffraction diagram {\it is\/} the indication of having
long-range positional order, and therefore of crystallinity. From the
point of view of condensed-matter physics, long-range positional order
is characterized by having Bragg peaks in the Fourier transform of the
density function $\rho(\rv)$. In the language of spectral theory, the
diffraction spectrum of $\rho(\rv)$ must contain a pure-point
component, for $\rho(\rv)$ to be considered crystalline.

\section{Periodic, quasiperiodic, and almost-periodic crystals}

Following H.~Bohr's theory of almost periodic functions~\cite{bohr},
solids whose density functions $\rho(\rv)$ may be expanded as a
superposition of a {\it countable\/} number of plane waves
\begin{equation}
  \label{eq:fourier}
  \rho(\rv) = \sum_{\kv\in L}\rho(\kv) e^{i\kv\cdot\rv},
\end{equation}
are to be called {\it almost periodic crystals}.  In particular, if
taking integral linear combinations of a finite number $D$ of wave
vectors in this expansion can span all the rest, then the crystal is
to be called {\it quasiperiodic.} The diffraction pattern of a
quasiperiodic crystal, therefore, contains Bragg peaks, each of which
can be indexed by $D$ integers.  If $D$ is the smallest number of wave
vectors that can span the whole set $L$ using integral linear
combinations, then $D$ is called the {\it rank,} or the {\it indexing
  dimension\/} of the crystal. {\it Periodic crystals\/} form a
special subset of all quasiperiodic crystals, whose rank $D$ is equal
to the actual physical dimension $d$ (the number of components in the
vectors $\rv$ and $\kv$).

For periodic crystals the set of Bragg peaks is truly discrete,
because the set of wave vectors $\kv$ in their Fourier expansion
(\ref{eq:fourier}) is discrete. For quasiperiodic crystals whose rank
$D$ is greater than the physical dimension $d$, the set $L$ of wave
vectors in the expansion (\ref{eq:fourier}) is dense---there are
$\kv$'s in $L$ that cannot be surrounded by a finite $d$-dimensional
ball that contains no other $\kv$'s. Nevertheless, in actual
experiments, where the total integrated diffraction intensity is
finite, Bragg peaks are not observed at wave vectors $\kv$ for which
the intensity $|\rho(\kv)|^2$ is below a certain threshold. In this
sense the observed diffraction pattern is essentially discrete,
abiding by the 1992 definition, even though the set $L$ is not
discrete. As the resolution of the diffraction experiment is improved,
additional peaks will appear in the diffraction diagram. If the
crystal is periodic the added peaks will appear further away from the
origin. If the crystal is quasiperiodic, new peaks can appear in
between already-existing ones, yet all the new peaks will still be
indexable by $D$ integers.  If the crystal is almost periodic, new
vectors may be required in order to generate the additional peaks,
thus the rank $D$ will seem to grow as the resolution is improved.

It should be noted that all experimentally observed crystals to date
are quasiperiodic. The term {\it quasicrystal}, which was first
introduced by Levine and Steinhardt~\cite{levine}, is simply short for
quasiperiodic crystal, but is used to refer to those quasiperiodic
crystals that are strictly aperiodic ({\it i.e.}\ with $D>d$). Some
authors require crystals to possess so-called ``forbidden symmetries''
in order to be regarded as quasicrystals. I have explained
elsewhere~\cite{quasidef} why such a requirement is inappropriate.

One should also recall that certain classes of quasiperiodic crystals
were known long before Shechtman's discovery. These are the so-called
{\it incommensurately-modulated crystals\/} and {\it incommensurate
  composite crystals,} (or {\it intergrowth compounds\/}).  These
crystals did not pose any serious challenge to the periodicity
paradigm because they could all be viewed as periodic structures that
had been slightly modified.  Order was still obtained through
periodicity, leaving the paradigm intact.

Almost periodic crystals have not (yet) been found in Nature, but can
be constructed artificially. A possible way of constructing such
crystals is to consider the limit of an infinite sequence of periodic
or quasiperiodic structures. If chosen properly, such a limit can
yield so-called {\it limit-periodic\/} or {\it limit-quasiperiodic\/}
structures, whose spectrum was shown to consist of Bragg peaks, even
though their rank is infinite~\cite{baakelimit}.

\section{What else is crystalline?}

The discovery of quasicrystals and the possibility for aperiodic order
has made a great impact on mathematics~\cite{senechal,moody,patera,%
  baakemoody}, as summarized by Lagarias~\cite{lagarias-order}. Much
effort has been invested in studying the characteristics of order, as
well as in the development of diffraction theory~\cite{hof,lagarias},
once it was realized that periodicity was not a necessary condition
for order and for the appearance of Bragg peaks in the diffraction
spectrum.

One interesting line of study~\cite{lagarias-order} is an attempt to
characterize the local mathematical properties of periodic crystals
and quasicrystals that are responsible for the emergence of long-range
order. For periodic crystals it is well-understood that any of a
number of local properties is sufficient for ensuring periodicity, and
thus order. These notions can be generalized to characterize the local
properties of quasicrystals, yet it is still unclear whether any of
these notions can serve as a sufficient condition for ensuring
quasiperiodicity, or more generally, long-range order.

Rather than trying to characterize local structural properties that
ensure Bragg peaks in the diffraction, a number of authors have
considered the diffraction properties of structural models that are
constructed by particular methods. Could there be a constructive
procedure that perfectly encompasses all structures that possess
long-range order? The answer to this question is still unknown,
although a number of unintuitive and even surprising results have
already been obtained.

It is well-understood that structure models that are constructed by
the cut-and-project technique from finite dimensional
spaces~\cite{senechal} are all quasiperiodic, and therefore have a
Bragg spectrum and possess long-range order. The same holds for
structures that are generated using the dual grid method and its
various generalizations~\cite{debruijn,rhyner}. These construction
methods are therefore too limited to encompass all ordered structures.
On the other hand, it turns out that other approaches that are
commonly used to generate quasicrystal models are too general, often
producing structures that do not possess Bragg peaks in their
diffraction spectrum, and therefore are not crystalline.

A typical example of a construction method that is too general to be
used to define crystallinity is that of substitution rules---either of
letter sequences in one dimension, or of tiles in two or three
dimensions.  The famous Fibonacci sequence, produced by the
substitution rules $L\to LS$ and $S\to L$, produces a rank-2
quasicrystal. What may seem as a minor change to these substitution
rules may generate sequences that do not have a Bragg spectrum.
Models based on such sequences do not possess long-range order and are
therefore not crystals, even though they are constructed using the
same {\it deterministic\/} procedure that is used to generate many
quasicrystals. It has been understood for some time~\cite{gaehler}
that what determines whether a structure, generated by substitution,
has long-range order is an algebraic property of the matrix, which
describes the substitution. The generated structure contains Bragg
peaks in its Fourier spectrum only if the largest eigenvalue of the
substitution matrix is a so-called Pisot-Vijayaraghavan number.

The fact that there are no correlations between the letters of a
sequence that is constructed by a deterministic procedure may seem
counter-intuitive, but readers who are familiar with the notion of
deterministic chaos may not be so alarmed.  Ben-Abraham~\cite{siba}
has suggested considering all structures that are generated by
deterministic procedures as crystals.  Along the same lines,
Mackay~\cite{mackay} has argued that
\begin{quote}{\it ``a crystal is a structure the description of which
    is much smaller than the structure itself.''}
\end{quote}
I object to such definitions that confuse the notion of order with
determinism. As counter-intuitive as it may seem,
deterministically-generated structures might be disordered. If this
were not the case it would not have been possible to design computer
algorithms that generate random numbers. Random number generators use
deterministic procedures to generate uncorrelated numbers that look as
if they were obtained by the random flipping of a coin. When
considering the actual structures, deterministic disorder is
practically the same as random disorder, and should not be associated
with crystallinity.  Not every deterministic structure that is simple
to generate is necessarily a crystal.

Baake and several co-workers~\cite{baakeentropy,baakevisible,%
  baakerandom,baakediffuse,baakechapter} are currently performing a
systematic study whose purpose is to characterize which distributions
of matter diffract to produce a pure point component in their
spectrum, and thus can qualify as possessing long-range order. In some
cases it is even difficult to determine whether the Fourier transform
of a structure exists, in the sense that it has a unique infinite
volume limit. When it is clear that the Fourier transform exists it
may still be difficult to rigorously know whether it contains Bragg
peaks. This is a difficult study which might take a long time to
complete, yet it is an important study that may eventually provide a
proper characterization of long-range order.

\section{The gray zone}

There are many structures that are partially-ordered---either by
design, or due to ``practical'' circumstances such as thermal
fluctuations, or the existence of defects and grain boundaries. In
some of these cases it might be difficult to decide at what point the
structure ceases to be crystalline. Often, details regarding the shape
of Bragg peaks offer a convenient way of doing so.  Melting of a
crystal due to thermal fluctuations, for example, may be formulated in
terms of a real-space condition such as the Lindemann criterion, or as
a condition on the reduction of Bragg-peak intensities due to the
Debye-Waller factor. Finite-size issues can be formulated as a
criterion of minimum size in real space, or as a condition on the
width of Bragg peaks or their scaling with the number of atoms. In
some of these cases it is a matter of taste to decide what constitutes
a crystal, and one may need to rely on common sense.  Questions such
as ``how many atoms are needed to form a crystal?'' or ``when is an
atomic cluster sufficiently large to be considered a nanocrystal?''
should not be of any real concern. On the other hand, questions of
short-range versus long-range order in a given homogeneous material
should be considered with care.

More interesting are structures that by design possess both long-range
order and some degree of disorder. An important example is that of
random-tiling models of quasicrystals~\cite{henley,baakerandom}. These
models are believed to have an underlying Bragg spectrum, indicative
of quasiperiodic long-range order, superimposed with a diffuse
background stemming from phason disorder, which is inherent to the
randomness of the models. In this case, the underlying Bragg spectrum
is often viewed as being the essential component, and therefore the
structure is considered to be crystalline. 

An opposite example is the one-dimensional Thue-Morse sequence,
generated from two letters by the substitution rule $L\to LS$ and $S\to
SL$. If represented as a sequence of short ($S$) and long ($L$)
segments with atoms at the joining points, then $1/2$ of the atoms
form a periodic lattice with a unit cell of length $S+L$. The
remaining atoms divide the unit cell into an $L$ followed by an $S$,
or {\it vice versa,} in a completely disordered fashion. In this case,
many authors tend to view the underlying periodic backbone as an
unimportant artifact, and the structure itself as being disordered. In
fact, one can carefully choose to represent the Thue-Morse sequence in
a non-generic way in which all of the Bragg peaks become extinguished,
leaving only the diffuse part of the diffraction
spectrum~\cite{gaehler}. If viewed as being essentially disordered,
the Thue-Morse sequence should not be regarded as a crystal, although
one should always remember that it does contain an underlying periodic
lattice.

\section{So, what is a crystal?}

Having said all this, I would like to conclude with a definition:
\begin{quote} 
\begin{enumerate}
\item A crystal is a solid that has long-range positional
  order.
\item Long-range positional order can be inferred from the
  existence of Bragg peaks in the Fourier spectrum of the solid.
\end{enumerate}
\end{quote}
It is common to add a descriptive prefix to the term crystal to refer
to soft materials, metamaterials, or any other form of matter that is
crystalline but is not an atomic solid. Examples are photonic crystals
and liquid or soft crystals---more appropriately called crystalline
liquids~\cite{wright} to distinguish them from the more common liquid
crystals that have long-range orientational rather than positional
order.  These should now be understood in the context of the new
definition of crystal to include periodic soft and photonic crystals
as well as soft quasicrystals~\cite{zeng,soft} and photonic
quasicrystals~\cite{jin,zoorob,frequency,defects} that are not
periodic.

I would happily welcome a revision of the second part of the
definition once a more complete characterization of order is
established. In the meantime, one must follow the only reliable
characterization in terms of the Fourier spectrum. I would be even 
happier to welcome a second scientific revolution in materials science
that would require yet another paradigm shift and a completely new
definition. Until that happens, I suggest sticking to order.

\section{Acknowledgments}

I thank S.I. Ben-Abraham for insisting that we reconsider now the
definition of crystal, rather than wait for another 10 or 15 years. I
also thank the Editor of {\it Zeit. Krist.}, Walter Steurer, for
suggesting to me that I put my thoughts on this matter into writing.
Finally, I thank David Mermin for his thoughtful comments on this
manuscript.

My research in crystallography is currently supported by the Israel
Science Foundation through grant number 684/06.



\end{document}